\documentclass{llncs}
\usepackage{llncsdoc}

\usepackage{amsmath}
\usepackage{amssymb}
\usepackage{xspace}
\usepackage[plain,Figure]{algorithm}
\usepackage[noend]{algpseudocode}
\usepackage{graphicx}
\usepackage{stmaryrd}
\usepackage{mathtools}
\usepackage{color}
\usepackage{comment}
\usepackage{multirow}
\usepackage{array}
\usepackage{booktabs}
\usepackage{pifont}
\usepackage{bbold}
\usepackage{wrapfig}
\usepackage{hyperref}
\usepackage{soul}
\usepackage{caption}
\usepackage{esvect}
\usepackage{algorithm}

\usepackage{bigstrut}

\begin{document}

\title{Learning  Abstractions for Program Synthesis}

\author{Xinyu Wang\inst{1}, Greg Anderson\inst{1}, Isil Dillig\inst{1} \and K. L. McMillan\inst{2}}
\institute{University of Texas, Austin \and Microsoft Research, Redmond}

\maketitle

\newcommand{\blaze}{\textsc{Blaze}\xspace}
\newcommand{\morpheus}{\textsc{Morpheus}\xspace}
\newcommand{\atlas}{\textsc{Atlas}\xspace}

\newcommand{\spec}{\varphi}
\newcommand{\trans}{\mathcal{T}}
\newcommand{\abs}{\mathcal{A}}
\newcommand{\prog}{\mathcal{P}}

\newcommand{\figref}[1]{Fig.~\ref{fig:#1}}
\newcommand{\figlabel}[1]{\label{fig:#1}}
\newcommand{\exref}[1]{Example~\ref{ex:#1}}
\newcommand{\exlabel}[1]{\label{ex:#1}}
\newcommand{\eqnref}[1]{Eqn.~\ref{eqn:#1}}
\newcommand{\eqnlabel}[1]{\label{eqn:#1}}

%%%% DO NOT CHANGE THESE %%%%%%%%%%%%%%
\renewcommand{\dots}{\cdot\cdot}
\renewcommand{\ldots}{\dots}
\renewcommand{\cdots}{\dots}
%%%%%%%%%%%%%%%%%%%%%%%%%%%%%%%%%%%%%%%%

\newcommand{\irule}[2]{\mkern-2mu\displaystyle\frac{#1}{\vphantom{,}#2}\mkern-2mu}

\newcommand{\todo}[1]{{\color{red}{#1}}}
\newcommand{\assign}{=}

\newcommand{\examples}{\mathcal{E}}

\newcommand{\grammar}{\mathcal{G}}
\newcommand{\dsl}{\mathcal{L}}
\newcommand{\inex}{{e_{\emph{in}}}}
\newcommand{\outex}{e_{\emph{out}}}
\newcommand{\cout}{{\vec{c}^{\prime}}}
\newcommand{\coutj}{{\vec{c}^{\prime}_{j}}}
\newcommand{\cin}{\vec{c}}
\newcommand{\atlasblaze}{{\sc Blaze}$^\star$}
\newcommand{\manualblaze}{{\sc Blaze}$^\dagger$}

\algnewcommand\Input{\textbf{input: }} 
\algnewcommand\Output{\textbf{output: }}

\newcommand{\tset}{\mathcal{I}}
\newcommand{\abstractions}{\mathcal{A}}
\newcommand{\transformers}{\mathcal{T}}
\newcommand{\absval}{\varphi}

\newcommand{\semantics}[1]{\llbracket{#1}\rrbracket}
\newcommand{\asemantics}[1]{{\semantics{#1}^\sharp}}
\newcommand{\ospec}{\Phi}

\newcommand{\itp}{\mathcal{I}}

\newcommand{\pre}{\spec_{pre}}
\newcommand{\post}{\spec_{post}}

\newcommand{\impr}{\times}

\vspace{-0.2in}
\begin{abstract}

Many example-guided program synthesis techniques use \emph{abstractions} to prune the search space. While  abstraction-based synthesis  has proven to be very powerful, a domain expert needs to provide a suitable abstract domain, together with the abstract transformers of each DSL construct. However,  coming up with useful abstractions can be non-trivial, as it requires both domain expertise and knowledge about the synthesizer. In this paper, we propose a new technique for learning abstractions that are useful for instantiating a general synthesis framework in a new domain. Given a DSL and a small set of training problems, our method uses \emph{tree interpolation} to infer reusable predicate templates that speed up synthesis in a given domain. Our method also learns suitable abstract transformers by solving a certain kind of second-order constraint solving problem in a data-driven way. We have implemented the proposed method in a tool called \atlas and evaluate it in the context of the \blaze meta-synthesizer. Our evaluation shows that (a) \atlas can learn useful abstract domains and transformers from few training problems, and (b) the abstractions learned by \atlas allow \blaze to achieve significantly better results compared to manually-crafted abstractions.

\end{abstract}

\vspace{-20pt}
\section{Introduction}
\vspace{-5pt}

Program synthesis is a powerful technique for automatically generating programs from high-level specifications, such as input-output examples. Due to its myriad use cases across a wide range of application domains (e.g., spreadsheet automation~\cite{flashfill,singh2016,fidex}, data science~\cite{morpheus,dace,mitra}, cryptography~\cite{synudic,dual},  improving programming productivity~\cite{sypet,insynth,prospector}), program synthesis has received widespread attention from the research community in recent years.

Because program synthesis is, in essence, a very difficult search problem, many recent solutions  prune the search space by utilizing \emph{program abstractions}~\cite{lambda2,morpheus,synquid,blaze,simpl,scythe}.  For example, state-of-the-art synthesis tools, such as \blaze~\cite{blaze}, \morpheus~\cite{morpheus} and Scythe~\cite{scythe}, symbolically execute (partial) programs over some abstract domain and reject those programs whose abstract behavior is inconsistent with the given specification. Because many programs share the same  behavior in terms of their abstract semantics, the use of abstractions allows these synthesis tools to significantly reduce the search space.

While the abstraction-guided synthesis paradigm has proven to be quite powerful, a down-side of such techniques is that they require a domain expert to manually come up with a suitable abstract domain and write abstract transformers for each DSL construct. For instance, the \blaze synthesis framework~\cite{blaze} expects a domain expert to manually specify a universe of predicate templates, together with sound abstract transformers  for every DSL construct.
%As another example, \morpheus also requires an abstract domain in the form predicate templates as well as the abstract semantics of every library method that can be used by the synthesizer. 
Unfortunately, this process is not only time-consuming but also requires significant insight about the application domain as well as the internal workings of the synthesizer.

In this paper, we propose a novel technique for automatically learning domain-specific abstractions that are useful for instantiating an example-guided synthesis framework in a new domain. Given a DSL and a training set of synthesis problems (i.e., input-output examples), our method learns  a useful abstract domain in the form of predicate templates and infers sound abstract transformers for each DSL construct. In addition to eliminating the significant manual effort required from a domain expert, the abstractions learned by our method often outperform manually-crafted ones in terms of their benefit to synthesizer performance.

The workflow of our approach, henceforth called \atlas~\footnote{\atlas stands
  for \underline{A}u\underline{T}omated \underline{L}earning of
  \underline{A}b\underline{S}tractions.}, is shown schematically in
  \figref{workflow}. Since \atlas is meant to be used as an \emph{off-line}
  training step for a general-purpose programming-by-example (PBE) system, it
  takes as input a DSL as well as a set of synthesis problems $\vec{\examples}$
  that can be used for training purposes. Given these inputs, our method enters
  a refinement loop where an \emph{Abstraction Learner} component discovers
  a sequence of increasingly precise abstract domains $\abs_1, \ldots, \abs_n$, and their corresponding abstract transformers $\trans_1, \ldots, \trans_n$, in order to help the \emph{Abstraction-Guided Synthesizer} (AGS)  solve all training problems. While the AGS can reject many incorrect solutions using an abstract domain $\abs_i$, it may still return some incorrect solutions due to the insufficiency of $\abs_i$. Thus, whenever the AGS returns an incorrect solution to {any} training problem, the Abstraction Learner discovers a more precise abstract domain and automatically synthesizes the corresponding abstract transformers. % such that the incorrect solutions can be refuted. 
  Upon termination of the algorithm, the final abstract domain $\abs_n$ and transformers $\trans_n$ are sufficient for the AGS to correctly solve \emph{all} training problems. Furthermore, because our method learns \emph{general} abstractions in the form of predicate templates, the learnt abstractions are expected to be useful for solving many \emph{other} synthesis problems beyond those in the training set.

\begin{figure}[t]
\vspace{-15pt}
\begin{center}
\includegraphics[scale=0.3]{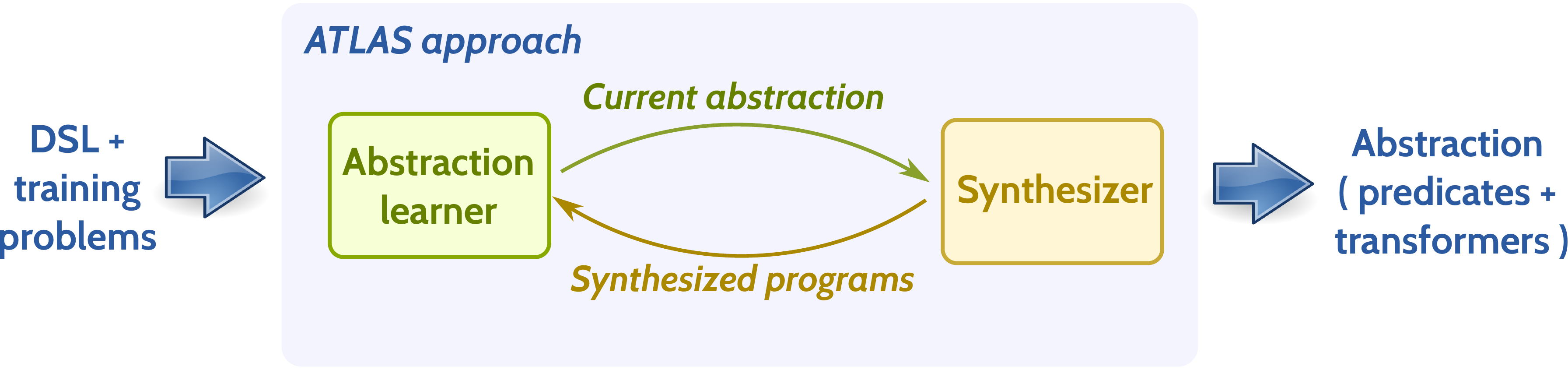}
\end{center}
\vspace{-15pt}
\caption{Schematic overview of our approach.}
\figlabel{workflow}
\vspace{-15pt}
\end{figure}

From a technical perspective, the Abstraction Learner uses two key ideas, namely \emph{tree interpolation} and \emph{data-driven constraint solving}, for learning useful abstract domains and  transformers respectively. Specifically, given an incorrect program $\prog$ that cannot be refuted by the AGS using the current abstract domain $\abs_i$, the Abstraction Learner generates a tree interpolant $\itp_i$ that serves as a proof of $\prog$'s incorrectness and constructs a new abstract domain $\abs_{i+1}$ by extracting templates from the predicates used in $\itp_i$. The Abstraction Learner also synthesizes the corresponding abstract transformers for $\abs_{i+1}$ by setting up a \emph{second-order constraint solving} problem where the goal is to find the unknown relationship between symbolic constants used in the predicate templates. Our method solves this problem in a data-driven way by sampling input-output examples for DSL operators and ultimately reduces the transformer learning problem to solving a system of linear equations.
%%However, because the resulting abductive reasoning problem is difficult to solve using existing techniques, we solve this problem in a data-driven way by leveraging linearity assumptions. 

We have implemented these ideas in a tool called \atlas and evaluate it in the context of the  \blaze program synthesis framework~\cite{blaze}.  Our evaluation shows that the proposed technique eliminates the manual effort involved in designing useful abstractions. More surprisingly, our evaluation also shows that the abstractions generated by \atlas outperform manually-crafted ones in terms of the performance of the \blaze synthesizer in two different application domains. 

To summarize, this paper makes the following key contributions: 
\vspace{-3pt}
\begin{itemize}
\item 
We describe a method for learning abstractions (domains/transformers) that are useful for instantiating program synthesis frameworks in new domains.
\vspace{0pt}
\item 
We show how tree interpolation can be used for learning abstract domains (i.e., predicate templates) from a few training problems.
\vspace{0pt}
\item 
We describe a method for automatically synthesizing transformers for a given abstract domain under certain assumptions. Our method is guaranteed to find the unique best transformer if one exists.
\vspace{0pt}
\item 
We implement our method in a tool called \atlas and experimentally evaluate it in the context of the \blaze synthesis framework. Our results demonstrate that the abstractions discovered by \atlas outperform manually-written ones used for evaluating \blaze in two application domains. 
\end{itemize}

\vspace{-15pt}
\section{Illustrative Example}\label{sec:example}
\vspace{-5pt}

Suppose that we wish to use  the \blaze meta-synthesizer to automate the class of string transformations considered by FlashFill~\cite{flashfill} and BlinkFill~\cite{blinkfill}. In the original version of the \blaze framework,  a domain expert needs to come up with a universe of suitable predicate templates as well as abstract transformers for each DSL construct~\cite{blaze}. We will now illustrate how \atlas automates this process, given a suitable DSL  and its semantics (e.g., the one used in~\cite{blinkfill}).

In order to use \atlas, one needs to provide a set of synthesis problems $\vec{\examples}$ (i.e., input-output examples) that will be used in the training process. Specifically, let us consider the three synthesis problems given below:
\vspace{-5pt}
\[
\small 
\vec{\examples} = \left\{ \ 
\begin{array}{lll}
\vspace{3pt}
\examples_1 & : & \big\{ \ ``\texttt{CAV}" \mapsto ``\texttt{CAV2018}"  , ``\texttt{SAS}" \mapsto ``\texttt{SAS2018}"   ,  ``\texttt{FSE}" \mapsto ``\texttt{FSE2018}"   \  \big\},   \\ 
\vspace{2pt}
\examples_2 & : & \big\{ \ ``\texttt{510.220.5586}" \mapsto ``\texttt{510-220-5586}" \ \big\}, \\ 
\examples_3 & : & \left\{ 
\begin{array}{ll}
``\texttt{$\backslash$Company$\backslash$Code$\backslash$index.html}" \mapsto ``\texttt{$\backslash$Company$\backslash$Code$\backslash$}" ,   \\ 
``\texttt{$\backslash$Company$\backslash$Docs$\backslash$Spec$\backslash$specs.html}" \mapsto ``\texttt{$\backslash$Company$\backslash$Docs$\backslash$Spec$\backslash$}"  \\ 
\end{array}
\right\}
\end{array}
\ \right\}.
\vspace{-3pt}
\]

%We will now illustrate how \atlas uses these training examples to learn a useful abstract domain $\abs$ and the corresponding transformers $\trans$. 

In order to construct the abstract domain $\abs$ and transformers $\trans$, \atlas starts with the trivial abstract domain $\abs_0 = \{ \top \}$ and transformers $\trans_0$, defined as
$
\asemantics{F(\top, \ldots, \top)} \assign \top
$
for each DSL construct $F$. Using this abstraction, \atlas invokes \blaze to find a program $\prog_0$ that satisfies specification $\examples_1$ under the current abstraction $(\abs_0, \trans_0)$. However, since the program $\prog_0$ returned by \blaze is incorrect with respect to the concrete semantics, \atlas tries to find a more precise abstraction that allows \blaze to succeed.

Towards this goal, \atlas enters a refinement loop that culminates in the discovery of the abstract domain $\abs_1 = \{ \top, \emph{len}(\fbox{$\alpha$}) = \texttt{c}, \emph{len}(\fbox{$\alpha$}) \neq \texttt{c} \}$, where $\alpha$ denotes a variable and {\tt c} is an integer constant. In other words, $\abs_1$ tracks equality and inequality constraints on the length of strings. After learning these predicate templates, \atlas also synthesizes the corresponding abstract transformers $\trans_1$. In particular, for each DSL construct, \atlas learns one abstract transformer for each combination of predicate templates used in $\abs_1$. For instance, for the {\tt Concat} operator which returns the concatenation $y$ of two strings $x_1, x_2$, \atlas synthesizes the following abstract transformers, where $\star$ denotes any predicate:
\vspace{-3pt}
\[\small 
\trans_1 = \left\{ \ \ 
\begin{array}{rcl}
\asemantics{\texttt{Concat} ( \top, \star) \big) } & \assign & \top \\
\asemantics{\texttt{Concat} ( \star, \top) \big) } & \assign & \top \\
\asemantics{\texttt{Concat} \big( \emph{len}(x_1)  \neq \texttt{c}_1, \emph{len}(x_2) \neq \texttt{c}_2 \big) } & \assign &  \top \\
\asemantics{\texttt{Concat} \big( \emph{len}(x_1)  =  \texttt{c}_1, \emph{len}(x_2) = \texttt{c}_2 \big) } & \assign &  \big( \emph{len}(y) = \texttt{c}_1 + \texttt{c}_2 \big) \\
\asemantics{\texttt{Concat} \big( \emph{len}(x_1)  =  \texttt{c}_1, \emph{len}(x_2) \neq \texttt{c}_2 \big) } & \assign &  \big( \emph{len}(y) \neq \texttt{c}_1 + \texttt{c}_2 \big) \\
\asemantics{\texttt{Concat} \big( \emph{len}(x_1)  \neq \texttt{c}_1, \emph{len}(x_2) = \texttt{c}_2 \big) } & \assign &  \big( \emph{len}(y) \neq \texttt{c}_1 + \texttt{c}_2 \big) 
\end{array}
\ \ \right\}.
\vspace{-3pt}
\]
Since the AGS can successfully solve $\examples_1$ using $(\abs_1, \trans_1)$, \atlas now moves on to the next training problem. 

For synthesis problem $\examples_2$, the current abstraction $(\abs_1, \trans_1)$ is \emph{not} sufficient for \blaze to discover the correct program. After processing $\examples_2$, \atlas refines the abstract domain to the following set of predicate templates:
\vspace{-5pt}
\[\small
\abs_2 = 
\big\{  \ 
\top, \emph{len}(\fbox{$\alpha$}) = \texttt{c}, \emph{len}(\fbox{$\alpha$}) \neq \texttt{c}, \emph{charAt}(\fbox{$\alpha$}, \texttt{i}) = \texttt{c}, \emph{charAt}(\fbox{$\alpha$}, \texttt{i}) \neq \texttt{c} 
\ \big\}. 
\vspace{-5pt}
\]
Observe that \atlas has  discovered two additional  predicate templates that track positions of characters in the string. \atlas also learns the corresponding abstract transformers $\trans_2$ for $\abs_2$. 

Moving on to the final training problem $\examples_3$, \blaze can already successfully solve it using $(\abs_2, \trans_2)$; thus, \atlas terminates with this abstraction. 
%Using this abstraction, \blaze can actually outperform the one with manually-crafted abstract domain on the benchmarks used in the evaluation of \blaze~\cite{blaze}. 

\vspace{-5pt}
\section{Overall Abstraction Learning Algorithm}
\vspace{-5pt}

\begin{figure}[!t]
\vspace{-35pt}
\begin{algorithm}[H]
\begin{algorithmic}[1]
\small 
\Procedure{LearnAbstractions}{$\dsl, \vec{\examples}$} 
\vspace{3pt}
\Statex \Input{Domain-specific language $\dsl$ and a set of training problems $\vec{\examples}$.}
\Statex \Output{Abstract domain $\abs$ and transformers $\trans$.}
\vspace{3pt}
\State {$\abstractions \gets \big\{  \ \top \  \big\}$;} \Comment{Initialization.}
\State {$\trans \gets \big\{ \ \asemantics{F(\top, \ldots, \top)} \assign \top \ | \ F \in \textsf{Constructs}(\dsl) \ \big\}$;}
\vspace{3pt}
\For{$i \gets 1, \dots, |\vec{\examples}|$}  
\vspace{2pt}
\While{true}  \Comment{Refinement loop.}
\State {$\prog \gets \textsf{Synthesize} (\dsl, \examples_i, \abs, \trans)$;}  \Comment{Invoke AGS.}
\If{$\prog = null$} {\textbf{break};}  \EndIf  
\If{$\textsf{IsCorrect} (\prog, \examples_i)$} {\textbf{break};} \EndIf   
\State {$\abs \gets \abs \cup \textsc{LearnAbstractDomain} (\prog, \examples_i)$;} 
\State {$\trans \gets \textsc{LearnTransformers} (\dsl, \abs)$;} 
\EndWhile
\EndFor
\vspace{3pt}
\State \Return{$(\abs, \trans)$;}
\EndProcedure
\end{algorithmic}
\end{algorithm}
\vspace{-40pt}
\caption{Overall learning algorithm. $\textsf{Constructs}$ gives the DSL constructs in $\dsl$.}
\figlabel{overall}
\vspace{-10pt}
\end{figure}

Our top-level algorithm for learning abstractions, called  \textsc{LearnAbstractions}, is shown in \figref{overall}. The algorithm takes two inputs, namely a domain-specific language $\dsl$ (both syntax and semantics) as well as a set of training problems $\vec{\examples}$, where each problem is specified as a \emph{set} of input-output examples $\examples_i$. The output of our algorithm is a pair $(\abs, \trans)$, where $\abs$ is an abstract domain represented by a set of predicate templates and  $\trans$ is the corresponding abstract transformers.

At a high-level, the \textsc{LearnAbstractions}  procedure starts with the most imprecise abstraction (just consisting of $\top$) and incrementally improves the precision of the abstract domain $\abs$ whenever the AGS fails to synthesize the correct program using $\abs$. Specifically, the outer loop (lines 4--10) considers each training example $\examples_i$ and performs a fixed-point computation (lines 5--10) that terminates when the current abstract domain $\abs$ is good enough to solve training problem $\examples_i$. Thus, upon termination, the learnt abstract domain $\abs$ is sufficiently precise for the AGS to solve all training problems $\vec{\examples}$.

Specifically, in order to find an abstraction that is sufficient for solving $\examples_i$, our algorithm invokes the AGS with the current abstract domain $\abs$ and corresponding transformers $\trans$ (line 6). We assume that \textsf{Synthesize} returns a program $\prog$ that is consistent with $\examples_i$ under abstraction ($\abs$, $\trans$). That is, symbolically executing $\prog$ (according to $\trans$) on inputs $\examples_i^\emph{in}$  yields abstract values $\vec{\absval}$ that are consistent with the outputs $\examples_i^\emph{out}$ (i.e., $\forall j. \ \examples_{ij}^\emph{out}  \in \gamma(\absval_j)$). However, while $\prog$ is guaranteed to be consistent with $\examples_i$ under the abstract semantics, it may not satisfy $\examples_i$ under the concrete semantics. We refer to such a program $\prog$ as \emph{spurious}.

Thus, whenever the call to \textsf{IsCorrect} fails at line 8, we invoke the {\sc LearnAbstractDomain} procedure (line 9) to learn additional predicate templates that are later added to $\abs$. Since the refinement of $\abs$  necessitates the synthesis of new transformers, we then call {\sc LearnTransformers} (line 10) to learn a new $\trans$. The new abstraction is guaranteed to rule out the spurious program $\prog$ as long as  there is a unique best transformer of each DSL construct for domain $\abs$.
%These transformers are guaranteed to rule out the spurious program $\prog$ as long as  there is a unique best transformer of each DSL construct for domain $\abs$.

\vspace{-10pt}
\section{Learning Abstract Domain using Tree Interpolation}\label{sec:abstract}
\vspace{-5pt}

In this section, we present the {\sc LearnAbstractDomain} procedure: Given a spurious program $\prog$ and a synthesis problem $\examples$ that $\prog$ does not solve, our goal is to find new predicate templates $\abs^\prime$ to add to the abstract domain $\abs$ such that the Abstraction-Guided Synthesizer no longer returns $\prog$ as a valid solution to the synthesis problem $\examples$. Our key insight is that we can mine for such useful predicate templates by constructing a \emph{tree interpolation} problem. In what follows,  we first review tree interpolants (based on~\cite{treevampire}) and then explain how we use this concept to find useful predicate templates.

\begin{definition}[Tree interpolation problem]
A tree interpolation problem $T = (V, r, P, L)$ is a directed labeled tree, where $V$ is a finite set of nodes, $r \in V$ is the root, $P: (V \backslash \{ r \}) \mapsto V$ is a function that maps children nodes to their parents, and $L : V \mapsto \mathbb{F}$ is a labeling function that maps nodes to formulas from a set $\mathbb{F}$ of first-order formulas such that $\bigwedge_{v \in V} L(v)$ is unsatisfiable. 
\vspace{-2pt}
\end{definition}

In other words, a tree interpolation problem is defined by a tree $T$ where each node is labeled with a formula and the  conjunction of these formulas is unsatisfiable. In what follows, we write $\emph{Desc}(v)$ to denote the set of all descendants of node $v$, including $v$ itself, and we write $\emph{NonDesc}(v)$ to denote all nodes other than those in $\emph{Desc}(v)$ (i.e., $V\backslash\emph{Desc}(v)$). Also, given a set of nodes $V'$, we write $L(V')$ to denote the set of all formulas labeling nodes in $V'$.

Given a tree interpolation problem $T$, a \emph{tree interpolant} $\itp$ is an annotation from every node in $V$ to a formula such that the label of the root node is \emph{false} and the label of an internal node $v$ is entailed by the conjunction of annotations of its children nodes. More formally, a tree interpolant is defined as follows:

\vspace{-5pt}
\begin{definition}[Tree interpolant]~\label{def:tree-itp}
Given a tree interpolation problem $T = (V, r, P, L)$, a tree interpolant for $T$ is a function $\itp: V \mapsto \mathbb{F}$ that satisfies the following conditions: 
\vspace{-5pt}
\begin{enumerate} 
\item 
$\itp(r) = \emph{false}$; 
\item 
For each $v \in V$: $\Big( \big( \bigwedge_{P(c_i) = v} \itp(c_i) \big) \wedge L(v) \Big) \Rightarrow \itp(v)$; 
\item 
For each $v \in V$: $\textnormal{\emph{Vars}} \big( \itp(v) \big) \subseteq \textnormal{\emph{Vars}} \big( L( { \small {\textnormal{\emph{Desc}}}(v) }) \big) \bigcap \textnormal{\emph{Vars}} \big( L(  { \small \textnormal{\emph{NonDesc}}(v)  } ) \big)$. 
\end{enumerate}
\vspace{-10pt}
\end{definition}

\begin{wrapfigure}{h}{0.5\linewidth}
\centering
\vspace{-20pt}
\includegraphics[scale=.26]{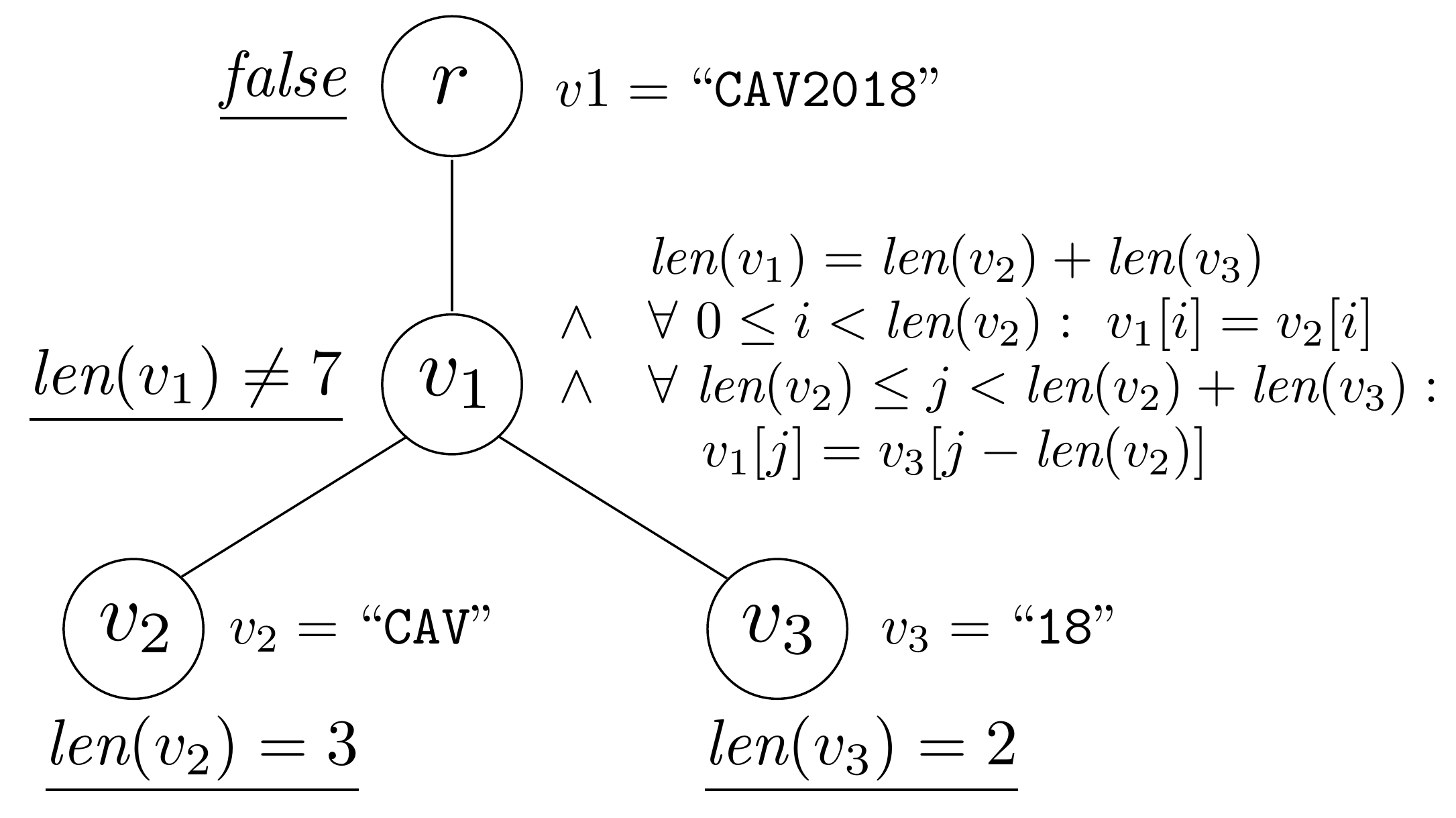}
\vspace{-3pt}
\caption{A tree interpolation problem and a tree interpolant (underlined).}
\figlabel{interpproblem}
\vspace{-20pt}
\end{wrapfigure}

\vspace{-2pt}
Intuitively, the first condition ensures that $\itp$ establishes the unsatisfiability of formulas in $T$, and the second condition states that $\itp$ is a valid annotation. As standard in Craig interpolation~\cite{itp1,itp2}, the third condition stipulates a ``shared vocabulary" condition by ensuring that the annotation at each node $v$ refers to the common variables between the descendants and non-descendants of $v$.

\vspace{-5pt}
\begin{example}
Consider the tree interpolation problem $T = (V, r, P, L)$ in \figref{interpproblem}, where $L(v)$ is shown to the right of each node $v$. 
A tree interpolant $\itp$ for this problem maps each node to the corresponding underlined formula. For instance, we have $\itp(v_1) = (\emph{len}(v_1) \neq 7)$. It is easy to confirm that $\itp$ is a valid interpolant according to Definition~\ref{def:tree-itp}. %Observe that $\itp$ satisfies all three conditions of Definition~\ref{def:tree-itp}.
\exlabel{interpproblem}
\vspace{-5pt}
\end{example}

To see how tree interpolation is useful for learning predicates, suppose that the spurious program $\prog$ is represented as an abstract syntax tree (AST), where each non-leaf node is labeled with the axiomatic semantics of the corresponding DSL construct. Now, since $\prog$ does not satisfy the given input-output example $(\inex, \outex)$, we can use this information to construct a labeled tree where the conjunction of labels is unsatisfiable. Our key idea is to mine useful predicate templates from  the formulas used in the resulting tree interpolant.

\begin{figure}[!t]
\vspace{-35pt}
\begin{algorithm}[H]
\begin{algorithmic}[1]
\Procedure{LearnAbstractDomain}{$\prog, \examples$} 
\vspace{0.05in}
\Statex \Input{Program $\prog$ that does not solve problem $\examples$ (set of examples).}
\Statex \Output{Set of predicate templates $\abs^\prime$.} 
\vspace{0.05in}
\State $\abs^\prime \gets \emptyset$; 
\For{\textbf{each} $(\inex, \outex) \in \examples$}
\If{$\semantics{\prog} \inex \neq \outex$}
\State $T \gets \textsc{ConstructTree}(\prog, \inex, \outex)$; 
\State $\itp \gets \textsf{FindTreeItp}(T)$; 
\For{\textbf{each} $v \in \textsf{Nodes}(T) \backslash \{ r \}$}
\State $\abs^\prime \gets \abs^\prime \cup \big\{ \textsf{MakeSymbolic} \big( \itp(v) \big) \big\}$;
\EndFor
\EndIf
\EndFor
\State \Return $\abs^\prime$; 
\EndProcedure
\end{algorithmic}
\end{algorithm}
\vspace{-40pt}
\caption{Algorithm for learning abstract domain using tree interpolation.}
\figlabel{learndomain}
\vspace{-10pt}
\end{figure}

With this intuition in mind, let us consider the {\sc LearnAbstractDomain} procedure shown in ~\figref{learndomain}: The algorithm uses a procedure called \textsc{ConstructTree} to generate a tree interpolation problem $T$ for each input-output example $(\inex, \outex)$\footnote{Without loss of generality, we assume that programs take a single input $x$, as we can always represent multiple inputs as a list.} that program $\prog$ does not satisfy (line 4). Specifically, letting $\Pi$ denote the AST representation of $\prog$, we construct $T=(V, r, P, L)$ as follows:
\vspace{-6pt}
\begin{itemize}
\item 
$V$ consists of all AST nodes in $\Pi$ as well as a ``dummy" node $d$. 
\item 
The root  $r$  of $T$ is the dummy node $d$. 
\item 
$P$ is a function that maps children AST nodes to their parents and maps the root AST node to the dummy node $d$. 
\item 
$L$ maps each node $v \in V$ to a formula as follows: 
\vspace{-5pt}
\[
\arraycolsep=1pt\def\arraystretch{1.4}
\small 
\centering 
L(v) = \left\{ 
\begin{array}{ll}
v^\prime = \outex  & v \text{ is the dummy root node with child } v^\prime.  \\ 
v = \inex &  v \text{ is a leaf representing program input.} \\  
v = c & v \text{ is a leaf representing constant } c. \\ 
\phi_{F}[\vec{v}^\prime / \vec{x}, v/y] \ \ & 
\arraycolsep=1pt\def\arraystretch{0.6}
\begin{array}{ll}
\\
v \text{ represents DSL operator } F \text{ with axiomatic semantics } \\ 
\phi_{F}(\vec{x}, y) \text{ and } \vec{v}^\prime \text{ represents children of } v.
\end{array}
\end{array}
\right.
\]
\end{itemize}

Essentially, the \textsc{ConstructTree} procedure labels any leaf node representing the program input with the input example $\inex$ and the root  node 
with the output example $\outex$.  All other internal nodes are labeled with the axiomatic semantics of the corresponding DSL operator (modulo renaming).\footnote{Here, we assume access to the  DSL's  axiomatic semantics. If this is not the case (i.e., we are only given the DSL's operational semantics), we can still annotate each node as $v = c$ where $c$ denotes the output of the partial program rooted at node $v$ when executed on $\inex$. However, this may affect the quality of the resulting interpolant.} Observe that the formula
$\bigwedge_{v\in V} L(v)$ is guaranteed to be unsatisfiable since $\prog$ does not satisfy the I/O example $(\inex, \outex)$; thus, we can obtain a tree interpolant for $T$.% using existing solvers, such as iZ3~\cite{iZ3}.

\begin{example}
\vspace{-5pt}
Consider program $\prog: \texttt{Concat}(x, ``\textsf{18}")$ which concatenates constant string ``\textsf{18}" to  input  $x$.  \figref{interpproblem} shows the result of invoking \textsc{ConstructTree} for $\prog$ and input-output example $(``\texttt{CAV}", ``\texttt{CAV2018}")$. As mentioned in \exref{interpproblem}, the tree interpolant $\itp$ for this problem is indicated with the underlined formulas. 
\vspace{-15pt}
\end{example}

Since the tree interpolant $\itp$ effectively establishes the incorrectness of program $\prog$, the predicates used in $\itp$ serve as useful abstract values that the synthesizer (AGS) should consider during the synthesis task. Towards this goal, the {\sc LearnAbstractDomain} algorithm iterates over each predicate used in $\itp$ (lines 7--8 in \figref{learndomain}) and converts it to a suitable template by replacing the constants and variables used in $\itp(v)$ with symbolic names (or ``holes"). Because the original predicates used in $\itp$ may be too specific for the current input-output example, extracting templates from the interpolant allows our method to learn reusable abstractions.

\begin{example}
\vspace{-5pt}
Given the tree interpolant $\itp$  from \exref{interpproblem},  \textsc{LearnAbstractDomain}   extracts two predicate templates, namely, $\emph{len}(\fbox{$\alpha$}) = \textsf{c}$ and $\emph{len}(\fbox{$\alpha$}) \neq \textsf{c}$. 
\vspace{-5pt}
\end{example}

\vspace{-25pt}
\section{Synthesis of Abstract Transformers}\label{sec:transformer}
\vspace{-5pt}

In this section, we turn our attention to the {\sc LearnTransformers} procedure for synthesizing abstract transformers $\trans$ for a given abstract domain $\abs$.
Following  presentation in prior work~\cite{blaze}, we consider abstract transformers that are described using equations of the following form:
\vspace{-5pt}
\begin{equation}
\asemantics{F \big( \chi_1(x_1, \vec{c}_1), \ldots, \chi_n(x_n, \vec{c}_n) \big)} 
= 
\bigwedge_{1 \leq j \leq m} {\chi'_{j} \big( y, \vec{f}_j(\cin) \big)}
\eqnlabel{trans}
\vspace{-8pt}
\end{equation}
%Here, $F$ is a DSL construct, $\chi_i, \chi'_{j}$ are predicate templates~\footnote{We assume that $\chi_1^{\prime}, \dots, \chi_m^{\prime}$ are distinct.}, $x_i$ is the $i$'th input of $F$, $y$ is $F$'s output, $\vec{c_1}, \ldots, \vec{c_n}$ are vectors of \emph{symbolic} constants, and $\vec{f}_j$ denotes a vector of \emph{affine functions} over $\cin = \vec{c_1}, \dots, \vec{c_n}$. \textcolor{red}{Intuitively, given concrete predicates describing the inputs to $F$, the transformer returns concrete predicates describing the output.}  Given such a transformer $\tau$, we write $(\chi_j', \vec{f}_j) \in \mathsf{Outputs}(\tau)$ where $\tau$ is given in \eqnref{trans}.
Here, $F$ is a DSL construct, $\chi_i, \chi'_{j}$ are predicate templates~\footnote{We assume that $\chi_1^{\prime}, \dots, \chi_m^{\prime}$ are distinct.}, $x_i$ is the $i$'th input of $F$, $y$ is $F$'s output, $\vec{c_1}, \ldots, \vec{c_n}$ are vectors of \emph{symbolic} constants, and $\vec{f}_j$ denotes a vector of \emph{affine functions} over $\cin = \vec{c_1}, \dots, \vec{c_n}$. 
Intuitively, given concrete predicates describing the inputs to $F$, the transformer returns concrete predicates describing the output. 
Given such a transformer $\tau$, let $\mathsf{Outputs}(\tau)$ be the set of pairs $(\chi_j', \vec{f}_j)$ in \eqnref{trans}.

We define the soundness of a transformer $\tau$ for DSL operator $F$ with respect to $F$'s axiomatic semantics $\phi_F$. In particular, we say that the abstract transformer from~\eqnref{trans} is \emph{sound} if the following implication is valid: 
\vspace{-7pt}
\begin{equation} 
\Big ( \phi_{F}(\vec{x}, {y}) \land \bigwedge_{1 \leq i \leq n} \chi_i(x_i, \vec{c}_i) \Big ) 
\Rightarrow 
\bigwedge_{1 \leq j \leq m} {\chi'_{j} \big( y, \vec{f}_j(\cin) \big)}
\eqnlabel{sound-trans}
\vspace{-7pt}
\end{equation}
That is, the transformer for $F$ is sound if the (symbolic) output predicate is indeed implied by the (symbolic) input predicates according to $F$'s semantics.

\begin{comment}
Given a DSL construct $F$ with $n$ arguments and an abstract domain $\abs$ consisting of $m$ predicate templates, we synthesize \emph{sound} abstract transformers for $F$ with \emph{all} combinations of argument templates, by considering a total of $m^{n}$ candidate transformers of the shape given in \eqnref{trans}.\footnote{While the abstract values in \blaze~\cite{blaze} can be conjunctions of predicates, it suffices to consider abstract transformers of the shape given in \eqnref{trans} because conjunctions can be handled using the following sound but possibly imprecise transformer:
\[
\vspace{-5pt}
\scriptsize 
\asemantics{F \big( (\bigwedge_{i_1} {\chi_{i_1}} ), \dots, (\bigwedge_{i_n} {\chi_{i_n}}) \big)} = \bigsqcap_{i_1} \dots \bigsqcap_{i_n} \asemantics{{F(\chi_{i_1}, \dots, \chi_{i_n})}}
\]
}
\end{comment}

Our key observation is that the problem of  learning  sound transformers  can be reduced to solving the following \emph{second-order constraint solving} problem:
\vspace{-6pt}
\begin{equation} 
\exists \vec{f}. \ \forall \vec{V}.  {\Big (} \big ( \phi_{F}(\vec{x}, {y}) \land \bigwedge_{1 \leq i \leq n} \chi_i(x_i, \vec{c}_i) \big ) 
\Rightarrow 
\bigwedge_{1 \leq j \leq m} {\chi'_{j} \big( y, \vec{f}_j(\cin) \big) {\Big )} }
\eqnlabel{trans-inf}
\vspace{-6pt}
\end{equation}
where  $\vec{f} = \vec{f}_1, \dots, \vec{f}_m$ and $\vec{V}$ includes all variables and functions from \eqnref{sound-trans} other than $\vec{f}$. In other words, the goal of this constraint solving problem is to find interpretations of the unknown functions $\vec{f}$ that make \eqnref{sound-trans} valid. Our key insight is to solve this problem in a \emph{data-driven} way by exploiting the fact that each unknown function $f_{j,k}$ is affine. 

Towards this goal, we first express each affine function $f_{j,k}(\cin)$ as follows:
\vspace{-8pt}
\[
f_{j,k}(\cin) 
= 
p_{j,k,1} \cdot c_{1} + \dots + p_{j,k,|\cin|} \cdot c_{|\vec{c}|} + p_{j,k, |\cin| + 1}
\vspace{-5pt}
\]
where each $p_{j,k,l}$ corresponds to an unknown integer constant that we would like to learn. Now, arranging the coefficients of functions $f_{j,1}, \ldots, f_{j,|\vec{f}_j|}$ in $\vec{f}_j$ into a $|\vec{f}_j| \times (|\cin| + 1)$ matrix $P_j$, we can represent $\vec{f}_j(\vec{c})$ in the following way:
\vspace{-7pt}
\begin{equation}\small
\vec{f}_j(\vec{c})^{\intercal}
=
\underbrace{
\left[
\begin{array}{cccc}
f_{j,1}(\cin) \\ 
\dots \\ 
f_{j,|\vec{f}_j|}(\cin) \\ 
\end{array}
\right]
}_{\coutj^{\intercal}}
= 
\underbrace{
\left[
\begin{array}{cccc}
p_{j,1,1} & \dots & \ \ p_{j,1, |\cin| + 1} \\ 
\dots & & \dots \\ 
p_{j,|\vec{f}_j|,1} & \dots & \ \ p_{j,|\vec{f}_j|, |\cin| + 1} \\ 
\end{array}
\right]
}_{P_j}
\underbrace{
\left[
\begin{array}{c}
c_{1} \\ 
\dots \\ 
c_{|\cin|} \\ 
1 \\ 
\end{array}
\right]
}_{\cin^{\dagger}}
\eqnlabel{matrix}
\vspace{-10pt}
\end{equation}
where $\cin^\dagger$ is $\cin^{\intercal}$ appended with the constant $1$.
%~\footnote{If each function is linear rather than affine, we have $\cin^\star = \cin$.} 

Given this representation, it is easy to see that the problem of synthesizing the unknown functions $\vec{f}_1, \ldots, \vec{f}_m$ from ~\eqnref{sound-trans} boils down to finding the unknown matrices $P_1, \ldots, P_m$ such that each $P_j$ makes the following implication valid:
\vspace{-7pt}
\begin{equation} 
\Lambda \ \equiv \ 
\Big( 
 \Big ( (\coutj^{\intercal} = P_j \cin^\dagger ) \land \phi_{F}(\vec{x}, {y})
 \land \bigwedge_{1 \leq i \leq n} \chi_i(x_i, \vec{c}_i) \Big ) \Rightarrow \chi'_j(y, \coutj)
\Big) 
\eqnlabel{abduction}
\vspace{-8pt}
\end{equation}

Our key idea is to infer these unknown matrices $P_1, \ldots, P_m$ in a data-driven way by generating input-output examples of the form $[i_1, \ldots, i_{|\cin|}] \mapsto [o_1, \ldots, o_{|\vec{f}_j|}]$ for each $\vec{f}_j$. In other words, $\vec{i}$ and $\vec{o}$ correspond to instantiations of $\cin$ and $\vec{f}_j(\cin)$ respectively.  Given sufficiently many such examples for every $\vec{f}_j$, we can then reduce the problem of learning each unknown matrix $P_j$ to the problem of solving a system of linear equations.

Based on this intuition, the {\sc LearnTransformers} procedure from~\figref{learntransformer} describes our algorithm for learning abstract transformers $\trans$ for a given abstract domain $\abs$.
At a high-level, our algorithm synthesizes one abstract transformer for each DSL construct $F$ and $n$ argument predicate templates $\chi_1, \dots, \chi_n$. In particular, given $F$ and $\chi_1, \dots, \chi_n$, the algorithm constructs the ``return value" of the transformer as:
\[
\varphi = \bigwedge_{1 \leq j \leq m} \chi'_j(y, \vec{f}_j(\vec{c}))
\]
where $\vec{f}_j$ is the inferred affine function for each predicate template $\chi'_j$.

\begin{figure}[!t]
\vspace{-28pt}
\begin{algorithm}[H]
\begin{algorithmic}[1]
\Procedure{LearnTransformers}{$\dsl, \abs$}
\Statex \Input{DSL $\dsl$ and abstract domain $\abs$.}
\Statex \Output{A set of transformers $\trans$ for constructs in $\dsl$ and abstract domain $\abs$.}  
\vspace{5pt}
\For{\textbf{each} $F \in \textsf{Constructs}(\dsl)$}
\vspace{3pt}
\For{$( \chi_1, \dots, \chi_n ) \in \abs^{n}$}
\State $\varphi \gets \top$; \Comment{$\varphi$ is output of transformer.}
\vspace{3pt}
\For{$\chi'_j \in \abs$}
\vspace{3pt}
\State $E \gets \textsc{GenerateExamples}(\phi_F,  \chi'_j, \chi_1, \dots, \chi_n)$; 
\State $\vec{f}_j \gets \textsf{Solve}(E)$;
\vspace{3pt}
\If {$\vec{f}_j \neq null \wedge \mathsf{Valid}(\Lambda[\vec{f_j}])$}
\vspace{5pt}
$\varphi \gets (\varphi \wedge \chi'_j(y, \vec{f}_j(\vec{c}_1, \dots, \vec{c}_n)))$
\EndIf
\EndFor
\State $\trans \gets \trans \cup \big\{ \asemantics{F(\chi_1(x_1, \vec{c}_1), \dots, \chi_n(x_n, \vec{c}_n))} = \varphi \big\}$;
\EndFor
\EndFor
\State \Return $\trans$;
\EndProcedure 
\end{algorithmic}
\end{algorithm}
\vspace{-35pt}
\caption{Algorithm for synthesizing abstract transformers. $\phi_{F}$ at line 6 denotes the axiomatic semantics of DSL construct $F$. Formula $\Lambda$ at line 8 refers to \eqnref{abduction}.}
\figlabel{learntransformer}
\vspace{-10pt}
\end{figure}

The key part of our {\sc LearnTransformers} procedure is the inner loop (lines 5--8) for inferring each of these $\vec{f}_j$'s.
Specifically, given an output predicate template $\chi'_j$, our algorithm first generates  a set of input-output examples $E$ of the form $[p_1, \ldots, p_n] \mapsto p_0$ such that $
\asemantics{F(p_1 , \ldots, p_n)} = p_0 
$
is a sound (albeit overly specific) transformer. Essentially, each $p_i$ is a concrete instantiation of a predicate template, so the examples $E$ generated at line 6 of the algorithm can be viewed as sound input-output examples for the general symbolic transformer given in ~\eqnref{trans}. (We will describe the {\sc GenerateExamples} procedure in Section~\ref{sec:gen-ex}).

Once we generate these examples $E$, the next step of the algorithm is to learn
the unknown coefficients of matrix $P_j$ from \eqnref{abduction} by solving a
system of linear equations (line 7). Specifically, observe that we can use each
input-output example $[p_1, \ldots, p_n] \mapsto p_0$ in $E$ to construct one
row of \eqnref{matrix}. In particular, we can directly extract $\vec{c} =
\vec{c}_1, \ldots, \vec{c}_n$ from $p_1, \ldots, p_n$ and the corresponding
value of $\vec{f}_j(\vec{c})$ from $p_0$. Since we have one instantiation of
\eqnref{matrix} for each of the input-output examples in $E$, the problem of
inferring matrix $P_j$ now reduces to solving a system of linear equations
of the form $A P_j^T = B$ where $A$ is a $|E| \times (|\vec{c}| + 1)$ (input) matrix and
$B$ is a $|E| \times |\vec{f}_j|$ (output) matrix. Thus, a solution to the equation $A
P_j^T = B$ generated from $E$ corresponds to a candidate solution for matrix $P_j$, which in turn uniquely defines $\vec{f}_j$.

Observe that the call to \textsf{Solve} at line 7 may return \emph{null} if no affine function exists. Furthermore, any \emph{non-null} $\vec{f}_j$ returned by \textsf{Solve} is just a \emph{candidate} solution and may not satisfy \eqnref{abduction}. For example, this situation can arise if we do not have sufficiently many examples in  $E$ and end up discovering an affine function that is ``over-fitted" to the examples. Thus, the validity check at line 8 of the algorithm ensures that the learnt transformers are actually sound.

\vspace{-6pt}
\subsection{Example Generation}\label{sec:gen-ex}
\vspace{-5pt}

In our discussion so far, we assumed an oracle that is capable of generating valid input-output examples for a given transformer. We now explain our {\sc GenerateExamples} procedure from \figref{example-generation} that essentially implements this oracle. In a nutshell,  the goal of {\sc GenerateExamples} is to synthesize input-output examples of the form $[p_1, \ldots, p_n] \mapsto p_0$  such that $\asemantics{F(p_1, \ldots, p_n)} = p_0$ is sound where each $p_i$ is a concrete predicate (rather than symbolic).

Going into more detail,  {\sc GenerateExamples} takes as input the semantics $\phi_F$ of DSL construct $F$ for which we want to learn a transformer for as well as the input predicate templates $\chi_1, \ldots, \chi_n$ and output predicate template $\chi_0$ that are supposed to be used in the transformer. 
For any example $[p_1, \ldots, p_n] \mapsto p_0$ synthesized by {\sc GenerateExamples}, each concrete predicate $p_i$ is an instantiation of the predicate template $\chi_i$ where the symbolic constants used in $\chi_i$ are substituted with \emph{concrete} values.

%It is also worth noting that {\sc GenerateExamples} takes two additional inputs, namely a spurious program $\prog$ and the training examples $\examples$ that $\prog$ does not satisfy. Intuitively, these additional arguments are used to ensure that the learnt transformers are precise enough for proving the spuriousness of $\prog$ under the new abstraction.

\begin{figure}[!t]
\vspace{-38pt}
\begin{algorithm}[H]
\begin{algorithmic}[1]
\Procedure{GenerateExamples}{$\phi_F, \chi_0, \ldots, \chi_n$}
\vspace{0.05in}
\Statex \Input{Semantics $\phi_F$ of operator $F$ and templates $\chi_0, \dots, \chi_n$ for output and inputs.}
\Statex \Output{A set of valid input-output examples $E$ for DSL construct $F$.} 
\vspace{3pt}
\State $E \gets \emptyset$;  
\vspace{1pt}
\While{$\neg$\textsf{FullRank}($E$)}
\vspace{2pt}
\State Draw $(s_1, \dots, s_n)$ randomly from distribution $R_F$ over $\textsf{Domain}(F)$;
%\State $(s_1, \dots, s_n) \xleftarrow{R} \textsf{Domain}(F)$;  %\Comment{Randomly draw values for arguments.}
\State $s_0 \gets \semantics{F(s_1, \dots, s_n)}$;
%\Statex \Comment{Abstract each $s_i$}
\State $(A_0, \dots, A_n) \gets \textsf{Abstract}(s_0, \chi_0, \dots, s_n, \chi_n)$; 
\vspace{2pt}
\For{{\bf each} $(p_0, \dots, p_n) \in A_0 \times \dots \times A_n$}  
\If {$\textsf{Valid}  \big( \bigwedge_{1\leq i \leq n} p_i \wedge \phi_{F} \Rightarrow p_0  \big)$} %\Comment{Keep only valid examples.}
%\vspace{2pt}
$E \gets E \cup \big\{ [p_1, \dots, p_n] \mapsto p_0 \big\}$;
\EndIf
\EndFor
\EndWhile
\vspace{1pt}
\State \Return $E$; 
\EndProcedure 
\end{algorithmic}
\end{algorithm}
\vspace{-40pt}
\caption{Example generation for learning abstract transformers.}
\figlabel{example-generation}
\vspace{-10pt}
\end{figure}

Conceptually, the {\sc GenerateExamples} algorithm proceeds as follows: First, 
it generates \emph{concrete} input-output examples $[s_1, \ldots, s_n] \mapsto 
s_0$ by evaluating $F$ on randomly-generated inputs $s_1, \ldots, s_n$ (lines 
4--5). Now, for each concrete I/O example $[s_1, \ldots, s_n] \mapsto s_0$, we 
generate a set of \emph{abstract} I/O examples of the form $[p_1, \ldots, p_n] 
\mapsto p_0$ (line 6). Specifically, we assume that the return value $(A_0, 
\ldots, A_n)$ of \textsf{Abstract} at line 6 satisfies the following properties 
for every $p_i \in A_i$:
\vspace{-7pt}
\begin{itemize}
\item $p_i$ is an instantiation of template $\chi_i$. 
\item $p_i$ is a sound over-approximation of $s_i$ (i.e., $s_i \in \gamma(p_i)$). 
\item For any other $p_i'$ satisfying the above two conditions, $p_i'$ is not logically stronger than $p_i$. 
\vspace{-5pt}
\end{itemize}

In other words, we assume that \textsf{Abstract} returns a set of ``best" sound abstractions of $(s_0, \ldots, s_n)$ under predicate templates $(\chi_0, \ldots, \chi_n)$.

%Each abstract value (i.e., predicate) $p_i$ is chosen in such a way that (a) $p_i$ is an instantiation of $\chi_i$, and (b) $p_i$ is a sound over-approximation of $s_i$ (line 8). 

Next, given abstractions $(A_0, \ldots, A_n)$ for $(s_0, \ldots, s_n)$, we consider each candidate abstract example of the form $[p_1, \ldots, p_n] \mapsto p_0$ where $p_i \in A_i$. Even though each $p_i$ is a sound abstraction of $s_i$, the example $[p_1, \ldots, p_n] \mapsto p_0$ may not be valid according to the semantics of operator $F$. Thus, the validity check at line 8 ensures that each example added to $E$ is in fact valid.

%^Observe that the {\sc GenerateExamples} procedure adds examples to $E$ until the matrix representation of $E$ is \emph{full-rank} (line 4). Intuitively, the full-rank condition ensures that we can always find the correct affine function between $\cin$ and $\cout$ if one exists. Furthermore, the larger the number of examples that we add to $E$, the less likely it is that we encounter case (3) discussed above for the {\sc LearnTransformers} algorithm.
 
\vspace{-5pt}
\begin{example}
Given abstract domain $\abs = \{ \emph{len}(\fbox{$\alpha$}) = \textsf{c} \}$, 
suppose we want to learn an abstract transformer $\tau$ for the \texttt{Concat} operator of the following form:
\vspace{-4pt}
\[
\asemantics{\texttt{Concat} \big( \emph{len}(x_1) = \textsf{c}_1, \emph{len}(x_2) = \textsf{c}_2 \big)} 
= 
\big( \emph{len}(y) = f([\textsf{c}_1, \textsf{c}_2]) \big)
\vspace{-4pt}
\]

\definecolor{light-gray}{gray}{0.3}

We learn the affine function $f$ used in the transformer by first generating a set $E$ of I/O examples for $f$ (line 6 in \textsc{LearnTransformers}). 
In particular, \textsc{GenerateExamples} generates concrete input values for \textsf{Concat} at random and obtains the corresponding output values by executing \textsf{Concat} on the input values. For instance, it may generate $s_1 = ``abc"$ and $s_2 = ``de"$ as inputs, and obtain $s_0 = ``abcde"$ as output. Then, it abstracts these values under the given templates. In this case, we have an abstract example with $p_1 = \big(\emph{len}(x_1) = {{3}} \big)$, $p_2 = \big( \emph{len}(x_2) = {{2}} \big)$ and $p_0 = \big( \emph{len}(y) = {{5}} \big)$. 
Since $[p_1, p_2] \mapsto p_0$ is a valid example, it is added in $E$ (line 8 in \textsc{GenerateExamples}). 
At this point, $E$ is not yet full rank, so the algorithm keeps generating more examples. 
Suppose it generates two more valid examples 
$\big( \emph{len}(x_1) = {{1}}, \emph{len}(x_2) = {{4}} \big) \mapsto \big(
\emph{len}(y) = {{5}} \big)$ and 
$\big( \emph{len}(x_1) = 6, \emph{len}(x_2) = 4 \big) \mapsto \big( \emph{len}(y) = 10 \big)$. 
Now $E$ is full rank, so \textsc{LearnTransformers} computes $f$ by solving the following system of linear equations: 
\vspace{-6pt}
\[\small 
\left[
\begin{array}{ccc}
{3} & {2} & 1 \\ 
{1} & {4} & 1 \\ 
6 & 4 & 1 \\
\end{array}
\right]
P^T
=
\left[
\begin{array}{ccc}
{5} \\ 
{5} \\ 
10 \\
\end{array}
\right]
\xRightarrow{\textsf{ \ Solve \ }}{}
P = 
\left[
\begin{array}{ccc}
1 & 1 & 0 \\ 
\end{array}
\right]
\vspace{-2pt}
\]
Here, $P$ corresponds to the function $f([ \texttt{c}_1, \texttt{c}_2]) = \texttt{c}_1 + \texttt{c}_2$, and this function defines the sound transformer: 
$
\asemantics{\texttt{Concat} \big( \emph{len}({x_1}) = \texttt{c}_1, \emph{len}({x_2}) = \texttt{c}_2 \big)} = \big( \emph{len}({y}) = \texttt{c}_1 + \texttt{c}_2 \big) 
$
which is added to $\trans$ at line 9 in \textsc{LearnTransformers}.
\end{example}

\vspace{-6pt}
\section{Soundness and Completeness}\label{sec:theorems}
\vspace{-4pt}

In this section we present theorems stating some of the soundness, completeness, and termination guarantees of our approach. 
%All proofs can be found in the extended version of this paper~\cite{atlas_extended}. 
All the proofs can be found in the appendix.

\vspace{-1pt}
\begin{theorem}[Soundness of \textnormal{\textsc{LearnTransformers}}] 
\label{thm:trans-sound}
Let $\trans$ be the set of transformers returned by \textsc{LearnTransformers}. 
Then, every $\tau \in \trans$ is sound according to \eqnref{sound-trans}.
\vspace{-2pt}
\end{theorem}

The remaining theorems are predicated on the assumptions that for each DSL construct $F$ and input predicate templates $\chi_1, \dots, \chi_n$ (i) there exists a unique best abstract transformer and (ii) the strongest transformer expressible in \eqnref{sound-trans} is logically equivalent to the unique best transformer. Thus, before stating these theorems, we first state what we mean by a \emph{unique best abstract transformer}.

\vspace{-2pt}
\begin{definition}[Unique best function]
Consider a family of transformers of the shape
$
\asemantics{F \big( \chi_1(x_1, \vec{c}_1), \dots, \chi_n(x_n, \vec{c}_n) \big)} = \chi'(y, \star)
$.
We say that $\vec{f}$ is the unique best function for $(F, \chi_1, \ldots, \chi_n, \chi')$ if (a)   replacing $\star$ with $\vec{f}$ yields a sound transformer, and (b) replacing $\star$ with any other  $\vec{f'}$ yields a transformer that is either  unsound or strictly worse (i.e.,  $\chi'(y, \vec{f}) \Rightarrow \chi'(y, \vec{f}')$ and $\chi'(y, \vec{f}') \not\Rightarrow \chi'(y, \vec{f})$. )
%Let $F$ be a DSL construct, and let $\chi_1, \dots, \chi_n$ be the input templates and $\chi'$ be the output template. 
%We say that $\vec{f}$ is a unique best function for $F, \chi_1, \dots, \chi_n, \chi'$ if 
%(a) $\asemantics{F \big( \chi_1(x_1, \vec{c}_1), \dots, \chi_n(x_n, \vec{c}_n) \big)} = \chi'(y, \vec{f})$ is a sound abstract transformer, 
%and 
%(b) for any \emph{other} sound abstract transformer $\asemantics{F \big( \chi_1(x_1, \vec{c}_1), \dots, \chi_n(x_n, \vec{c}_n) \big)} = \chi'(y, \vec{f}')$ where $\vec{f} \neq \vec{f}'$, we have $\chi'(y, \vec{f}) \Rightarrow \chi'(y, \vec{f}')$ and $\chi'(y, \vec{f}') \not\Rightarrow \chi'(y, \vec{f})$. 
\vspace{-2pt}
\end{definition}

%In other words, $\vec{f}$ is the unique best function for a transformer of the form
%$
%\asemantics{F \big( \chi_1(x_1, \vec{c}_1), \dots, \chi_n(x_n, \vec{c}_n) \big)} = \chi'(y, \star)$
% if replacing $\star$ with $\vec{f}$ yields a sound transformer, but replacing it with any other  $\vec{f'}$ yields a transformer that is either  unsound or strictly worse.

We now define unique best transformer in terms of unique best function:

\vspace{-2pt}
\begin{definition}[Unique best transformer]
Let $F$ be a DSL construct and let $(\chi_1, \ldots, \chi_n) \in \abs^n$ be the input templates for $F$. 
We say that the abstract transformer $\tau$ is a unique best transformer for $F, \chi_1, \dots, \chi_n$ if 
%\vspace{-8pt}
%\[
%\asemantics{F \big( \chi_1(x_1, \vec{c}_1), \dots, \chi_n(x_n, \vec{c}_n) \big)} 
%= 
%\bigwedge_{1 \leq j \leq m} {\chi'_j (y, \vec{f}_j)}
%\vspace{-8pt}
%\]
%is a unique best abstract transformer for $F, \chi_1, \dots, \chi_n$ if 
(a) $\tau$ is sound, and 
(b) for any predicate template $\chi \in \abs$, we have $(\chi, \vec{f}) \in
\mathsf{Outputs}(\tau)$ if and only if $\vec{f}$ is a unique best function for $(F, \chi_1,
\ldots, \chi_n, \chi)$ for some affine $\vec{f}$. 
%for some $j \in [1, m]$ \emph{iff} $\vec{f}_j$ is a unique best function for $F, \chi_1, \dots, \chi_n, \chi$.} 
\vspace{-2pt}
\end{definition}
%\todo{XINYU: The $\leftarrow$ direction is actually wrong because two $\chi_j^{\prime}$'s might use the same $\vec{f}_j$. We cannot simply say $\chi'_j = \chi$. Maybe we should fall back to having three conditions?}

\vspace{-4pt}
\begin{definition}[Complete sampling oracle]
Let $F$ be a construct, $\abs$ an abstract domain, and $R_F$ a probability distribution over
$\textsc{Domain}(F)$ with finite support $S$. Futher, for any input predicate templates $\chi_1, \ldots, \chi_n$ and output predicate template $\chi_0$ in $\abs$ admitting a unique best function $\vec{f}$,
let $C(\chi_0,\ldots,\chi_n)$ be the set of tuples $(c_0,\ldots,c_n)$ such that $(\chi_0(y,c_0),\chi_1(x_1,c_1),\ldots,\chi_n(x_n,c_n)) \in A_0 \times \cdots \times A_n$ and $c_0 = \vec{f}(c_1,\ldots,c_n)$, where $A_0 \times \cdots \times A_n = \textsc{Abstract}(s_0,\chi_0,\ldots,s_n,\chi_n)$ and $(s_1,\ldots,s_n) \in S$ and $s_0 = \semantics{F(s_1, \dots, s_n)}$. 
The distribution $R_F$ is a \emph{complete sampling oracle} if $C(\chi_0,\ldots,\chi_n)$ has full rank for all
$\chi_0,\ldots,\chi_n$.
\vspace{-2pt}
\end{definition}

The following theorem states that {\sc LearnTransformers} is guaranteed to synthesize the best transformer if a unique one exists:

%%%%%% Greg's fix %%%%%%

\begin{comment}
\begin{theorem}[Completeness of \textnormal{\textsc{LearnTransformers}}]
\label{thm:trans-complete}
Let $\trans$ be the set of  transformers returned by \textsc{LearnTransformers}
for  abstract domain $\abs$. Let $\tau$ be the unique best transformer for a DSL
construct $F$ and input predicate templates $\chi_1  \dots, \chi_n \in \abs^n$.
Suppose there exist inputs $(s_1,\dots,s_n)\in\textsc{Domain}(F)$ such that for
$s_0 = \semantics{F(s_1, \dots, s_n)}$ there exists $(A_0, \dots, A_n)\in
\textsf{Abstract}(s_0, \dots, s_n)$, with $(A_0, \dots, A_n)$ corresponding to
$\tau$.  Then we have $\tau \in \trans$. 
\end{theorem}
\end{comment}

%%%%%%%%%%%%%%%%%%%%%%%%%%

%%%%%% Ken's fix %%%%%%

\vspace{-2pt}
\begin{theorem}[Completeness of \textnormal{\textsc{LearnTransformers}}]
\label{thm:trans-complete}
Given an abstract domain $\abs$ and a complete sampling oracle $R_F$ for $\abs$, \textsc{LearnTransformers} terminates.
Further, let $\trans$ be the set of transformers returned and let $\tau$ be the
unique best transformer for DSL
construct $F$ and input predicate templates $\chi_1, \dots, \chi_n \in \abs^n$.
Then we have $\tau \in \trans$. 
\vspace{-2pt}
\end{theorem}

%%%%%%%%%%%%%%%%%%%%%%%%%%

Using this completeness (modulo unique best transformer) result, we can now state the termination guarantees of our {\sc LearnAbstractions} algorithm:

\vspace{-5pt}
\begin{theorem}[Termination of \textnormal{\textsc{LearnAbstractions}}]
\label{thm:termination}
Given a complete sampling oracle $R_F$ for every abstract domain and the unique best transformer assumption, if there exists a solution for every problem $\examples_i \in \vec{\examples}$, then \textsc{LearnAbstractions} terminates. 
\vspace{-5pt}
\end{theorem}

\todo{
%Termination of our abstraction learning algorithm is predicated on two assumptions: that there exists a unique best transformer for each DSL construct for every abstract domain generated by the {\sc LearnAbstractDomain} procedure and that there exists a complete sampling oracle for these domains. These conditions may not hold in practice. However, we have found empirically that, for the string and
%matrix domains used in our evaluation, the obtained domains do have
%unique best transformers and that sampling uniformly over a small
%range of integer values is adequate to find them.
}

\vspace{-6pt}
\section{Implementation and Evaluation}
\vspace{-3pt}

We have implemented the proposed method as a new tool called \atlas, which is written in Java. 
\atlas takes as input a set of training problems,  an Abstraction-Guided Synthesizer (AGS), and a DSL
and returns an abstract domain (in the form of predicate templates) and the corresponding transformers. 
%\todo{Whenever a new abstraction $\abs$ is learnt, the AGS is run by taking $\abs$.} 
Internally, \atlas uses the Z3 theorem prover~\cite{z3} to compute tree interpolants and the JLinAlg linear algebra library~\cite{jlinalg} to solve linear equations.

To assess the usefulness of \atlas, we conduct an experimental evaluation in which our goal is to answer the following two questions:
\vspace{-6pt}
\begin{enumerate}
\item  How does \atlas perform during training? That is, how many training problems does it require and how long does training take?
\item How useful are the abstractions learnt by \atlas in the context of synthesis?
\vspace{-20pt}
\end{enumerate}

\vspace{-5pt}
\subsection{Abstraction Learning}
\vspace{-0pt}

To answer our first question, we use \atlas to automatically learn abstractions
for two application domains: (i) string manipulations and (ii) matrix
transformations. We provide \atlas with the DSLs used in~\cite{blaze} and employ
\blaze  as  the underlying Abstraction-Guided Synthesizer. 
Axiomatic semantics for each DSL construct were given in the theory of equality with uninterpreted functions. 
%as well as the  provide the DSLs used in \blaze for string and matrix transformations respectively. 

\vspace{4pt}
\noindent\textbf{\emph{Training set information.}}
For the string domain, our training set consists of exactly the  four problems used as motivating examples in the BlinkFill paper~\cite{blinkfill}. Specifically, each training problem consists of 4-6 examples that demonstrate the desired string transformation.
For the matrix domain, our training set consists of four (randomly selected) synthesis problems taken from online forums. Since almost all online posts contain a single input-output example, each training problem includes one example illustrating the desired matrix transformation. 

%\vspace{4pt}
%\noindent\textbf{\emph{Experimental setup.}}
%For both domains, we provide \atlas with the DSLs used for evaluating \blaze in the string and matrix domains~\cite{blaze} as well as the training problems described above. 
%In particular, each training problem $\examples_i$ terminates when the AGS either successfully solves $\examples_i$, proves that no DSL program can solve $\examples_i$, or times out in 10 minutes. 

\newcolumntype{?}{!{\vrule width .9pt}}

\renewcommand{\arraystretch}{1.2}

\begin{figure}[!t]
\vspace{-6pt}
\scriptsize  
\centering
\begin{minipage}{0.48\linewidth}
\centering
\begin{tabular}{c ? c c c | ccc | c }
\hline 
  & \multirow{2}{*}{ $|\abs|$} & \multirow{2}{*}{$|\trans|$} & \multirow{2}{*}{{Iters. }} & \multicolumn{4}{c}{Running time (sec)} \\ \cline{5-8}
  &  &  &  &  \ $T_{\textsf{AGS}}$ \ & $T_{\abs}$ \  & $T_{\trans}$  \  &  \ $T_{\emph{total}}$ \  \\  \hline\hline 
$\examples_1$ & 5 & 30 & 4 & 0.6 & 0.2 & 0.2 & 1.0  \\ 
$\examples_2$ & 5 & 30 & 1 & 4.9 & 0 & 0 & 4.9  \\ 
$\examples_3$ & 5 & 30 & 1 & 0.2 & 0 & 0  & 0.2 \\ 
$\examples_4$ & 5 & 30 & 1 & 4.1 & 0 & 0  & 4.1 \\ \hline\hline 
\text{Total}  & 5 & 30 & 7 & 9.8 & 0.2 & 0.2 & \textbf{10.2} \\ \hline 
\end{tabular}
\vspace{-7pt}
\caption*{{\small String domain}}
\end{minipage}
\ \ \ \ 
\begin{minipage}{0.48\linewidth}
\centering
\begin{tabular}{c ? c c c | ccc | c }
\hline 
  & \multirow{2}{*}{ $|\abs|$} &\multirow{2}{*}{$|\trans|$} & \multirow{2}{*}{{Iters. }} & \multicolumn{4}{c}{Running time (sec)} \\ \cline{5-8}
  &  &  &  &  \ $T_{\textsf{AGS}}$ \ & $T_{\abs}$ \  & $T_{\trans}$  \  &  \ $T_{\emph{total}}$ \  \\  \hline\hline 
$\examples_1$ & 8 & 45 & 3 & 2.9 & 0.7 & 0.5  & 4.1 \\ 
$\examples_2$ & 8 & 45 & 1 & 2.8 & 0 & 0 & 2.8 \\ 
$\examples_3$ & 10 & 59 & 2 & 0.5 & 0.3 & 0.2 & 1.0 \\ 
$\examples_4$ & 10 & 59 & 1 & 14.6 & 0 & 0 & 14.6 \\ \hline\hline 
\text{Total}  & 10 & 59 &  7 & 20.8 & 1.0 & 0.7 & \textbf{22.5} \\ \hline 
\end{tabular}
\vspace{-7pt}
\caption*{{\small Matrix domain}}
\end{minipage}
\vspace{-6pt}
\caption{Training results. 
$|\abs|, |\trans|$, {Iters} denote the number of predicate templates, abstract transformers, and iterations taken per training instance (lines 5-10 from \figref{overall}), respectively. 
$T_{\textsf{AGS}}, T_{\abs}, T_{\trans}$ denote the times for invoking the synthesizer (AGS), learning the abstract domain, and learning the abstract transformers, respectively. 
$T_{\emph{total}}$ shows the total training time in seconds.}
\figlabel{trainingresult}
\vspace{-2pt}
\end{figure}
\renewcommand{\arraystretch}{1}

\renewcommand{\arraystretch}{1}
\setlength{\extrarowheight}{0.5pt}

\begin{figure}[!t]
\vspace{-0pt}
\scriptsize   
\centering 
\begin{tabular}{ c ?   c c | c c |      c c | c c ? c | c c }
\hline 
& \multicolumn{4}{c|}{{\fontsize{7}{7}\selectfont {Original \manualblaze\ benchmarks}}} 
& \multicolumn{4}{c?}{{\fontsize{7}{7}\selectfont Additional benchmarks}} 
& \multicolumn{3}{c}{{\fontsize{7}{7}\selectfont {All benchmarks}}} \bigstrut 
\\ \cline{2-12} 
& \multicolumn{2}{c|}{\#Solved} 
& \multicolumn{2}{c|}{\begin{tabular}{@{}c@{}} \ {\fontsize{7}{7}\selectfont{Running time}} \ \\ {\fontsize{7}{7}\selectfont{improvement}} \end{tabular}} 
& \multicolumn{2}{c|}{\#Solved} 
& \multicolumn{2}{c?}{\begin{tabular}{@{}c@{}} \ {\fontsize{7}{7}\selectfont{Running time}}  \ \\ {\fontsize{7}{7}\selectfont{improvement}} \end{tabular}} 
& \multicolumn{1}{c|}{\begin{tabular}{@{}c@{}} \ {\fontsize{7}{7}\selectfont{Time}} \ \\ {\fontsize{7}{7}\selectfont{(sec)}} \end{tabular}} 
& \multicolumn{2}{c}{\begin{tabular}{@{}c@{}} \ {\fontsize{7}{7}\selectfont{Running time}} \ \\ {\fontsize{7}{7}\selectfont{improvement}} \end{tabular}} 
\\ \cline{2-12} 
& \ \atlasblaze & \manualblaze\  
& \  \ max. & avg.  
& \ \atlasblaze & \manualblaze\ 
& \ \  max. & avg.  
& avg. 
& \ \  max. & avg.  
\\ \hline \hline 
\textsf{String} \ 
& \ \textbf{93} &  91 \ 
& \ 15.7$\impr$ & 2.1$\impr$ \  
& \ \textbf{40} & 40 \  
& \ 56$\impr$ & 22.3$\impr$ \ 
& \textbf{2.8} 
& \ \textbf{56$\impr$} & \textbf{8.3$\impr$} 
\bigstrut\bigstrut 
\\  
\textsf{Matrix} \ 
& \ \textbf{39} & 39 \ 
& \ 6.1$\impr$ & 3.1$\impr$ \ 
& \ \textbf{20} & 19 \ 
& \ 83$\impr$ & 21.5$\impr$ 
& \textbf{5.0}  
&  \ \textbf{83$\impr$} & \textbf{9.2$\impr$}
\bigstrut\bigstrut 
\\ \hline  
\end{tabular}
\vspace{-6pt}
\caption{Improvement of \atlasblaze\ over \manualblaze\ on   string and matrix benchmarks. }
\figlabel{testresult}
\vspace{-12pt}
\end{figure}

\vspace{4pt}
\noindent\textbf{\emph{Main results.}}
Our main results are summarized in  \figref{trainingresult}. The main take-away message is that \atlas can learn abstractions quite efficiently and does not require a large training set. For example, \atlas learns 5 predicate templates and {30} abstract transformers for the string domain in a total of 10.2 seconds. Interestingly, \atlas does not need all the training problems to infer these four predicates and converges to the final abstraction after just processing the first training instance. 
Furthermore, for the first training instance, it takes \atlas 4 iterations in the learning loop (lines 5-10 from \figref{overall}) before it converges to the final abstraction. Since this abstraction is sufficient for solving the remaining training problems, the loop takes just one iteration.

Looking at the right side of \figref{trainingresult}, we also observe similar results for the matrix domain. In particular, \atlas learns 10 predicate templates and 59 abstract transformers in a total of 22.5 seconds.  Furthermore, \atlas converges to the final abstract domain after processing the first three problems~\footnote{The learnt abstractions can be found in the appendix.} and the number of iterations for each training instance is also quite small.

%\vspace{4pt}
%\noindent\textbf{\emph{Precision of abstract transformers.}}
%In general, the  algorithm presented in Section~\ref{sec:transformer} is not guaranteed to compute the strongest abstract transformer in the given domain. To evaluate the precision of our learnt transformers, we manually inspected each abstract transformer and found that \atlas learns precise transformers in all cases for both string and matrix domains. 

\vspace{-3pt}
\subsection{Evaluating the Usefulness of Learnt Abstractions}
\vspace{-1pt}

To answer our second question, we integrated the  abstractions synthesized by \atlas into the \blaze meta-synthesizer. In the remainder of this section, we refer to all instantiations of \blaze using the \atlas-generated abstractions as \atlasblaze. To assess how useful the  automatically generated abstractions are, we compare \atlasblaze\ against \manualblaze, which refers to the manually-constructed instantiations of \blaze described in~\cite{blaze}.

\vspace{2pt}
\noindent\textbf{\emph{Benchmark information.}}
For the string domain, our benchmark suite consists of (1) \emph{all} 108 string
transformation benchmarks that were used to evaluate \manualblaze\ and (2) 40 additional challenging problems that are collected from online forums which involve manipulating file paths, URLs, etc. The number of examples for each benchmark ranges from 1 to 400, with a median of 7 examples. 
For the matrix domain, our benchmark set includes (1) \emph{all} 39 matrix transformation benchmarks in the \manualblaze\ benchmark suite and (2) 20 additional challenging problems collected from online forums.
%Each benchmark comes with one input-output example, since almost all  online posts contain only a single example. 
\emph{We emphasize that the set of benchmarks used for evaluating \atlasblaze\ are completely \emph{disjoint} from the set of synthesis problems used for training \atlas.}

\vspace{2pt}
\noindent\textbf{\emph{Experimental setup.}} 
We evaluate \atlasblaze\ and \manualblaze\ using the same DSLs from the \blaze paper~\cite{blaze}. For each benchmark, we provide the same set of input-output examples to \atlasblaze\ and \manualblaze, and use a time limit of  20 minutes per synthesis task.

\vspace{2pt}
\noindent\textbf{\emph{Main results.}} 
Our main evaluation results are summarized in \figref{testresult}. The key observation is that \atlasblaze\ consistently improves upon \manualblaze\ for both string and matrix transformations.
In particular, \atlasblaze\ not only solves more benchmarks than \manualblaze\ for both domains, but also achieves about an order of magnitude speed-up on average for the common benchmarks that both tools can solve. 
Specifically, for the string domain, \atlasblaze\ solves 133 (out of 148) benchmarks within an average of 2.8 seconds and achieves an average  8.3$\impr$ speed-up over \manualblaze. 
%Note that the speed-up for the additional challenging problems (22.3$\impr$) is even more significant. 
%than that for the original benchmarks used for evaluating \manualblaze\ (2.1$\impr$). 
For the matrix domain, we also observe a very similar result where \atlasblaze\ leads to an overall speed-up of 9.2$\impr$ on average.

In summary, this experiment confirms that the abstractions discovered by \atlas are indeed useful and that they outperform manually-crafted abstractions despite eliminating  human effort.
%involved in instantiating a generic abstraction-guided synthesizer in a new domain.

\vspace{-13pt}
\section{Related Work}
\vspace{-10pt}

To our knowledge, this paper is the first one to
automatically learn abstract domains and transformers that are useful
for program synthesis. We also believe it is the first to apply
interpolation to program synthesis, although interpolation has been
used to synthesize other artifacts such as
circuits~\cite{DBLP:conf/fmcad/BloemEKKL14} and strategies for
infinite games~\cite{farzan2017strategy}. In what follows, we briefly
survey existing work related to program synthesis, abstraction
learning, and abstract transformer computations.

\vspace{-8pt}
\paragraph{\bf \em Program synthesis.} Our work is  intended to complement example-guided program synthesis techniques that utilize program abstractions to prune the search space~\cite{simpl,scythe,morpheus,blaze}. For example, {\sc Simpl}~\cite{simpl} uses abstract interpretation to speed up search-based synthesis and applies this technique to the generation of imperative programs for introductory programming assignments. Similarly, {\sc Scythe}~\cite{scythe} and  \morpheus~\cite{morpheus} perform enumeration over program {sketches} 
%(rather than complete programs) 
and use abstractions to reject  sketches that do not have any valid completion. 
Somewhat different from these techniques, \blaze constructs a finite tree automaton that accepts all programs whose behavior is consistent with the specification according to the DSL's abstract semantics. We believe that the method described in this paper can be useful to all such abstraction-guided synthesizers.

%Our work is most closely related to the \blaze synthesis framework~\cite{blaze} and is inspired directly from the observation that it is non-trivial for users to come up with domain-specific abstractions that are useful to the synthesizer. While the shape of the abstract transformers we learn are also inspired by \blaze, we believe that similar techniques could also be useful for learning other forms of transformers.

\vspace{-8pt}
\paragraph{\bf \em Abstraction refinement.}

In verification, as opposed to synthesis, there have been many
works that use Craig interpolants to refine
abstractions~\cite{itp2,blast1,ufo}. Typically,
these techniques generalize the interpolants to abstract domains by extracting a vocabulary of
predicates, but they do not generalize by adding parameters to
form templates. In our case, this is essential because interpolants
derived from fixed input values are too specific to be
directly useful. Moreover, we \emph{reuse} the resulting abstractions for
subsequent synthesis problems. In verification, this would be
analogous to re-using an abstraction from one property or program to
the next. It is conceivable that template-based generalization could
be applied in verification to facilitate such reuse.

\vspace{-10pt}
\paragraph{\bf \em Abstract transformers.}
Many verification techniques use logical abstract
domains~\cite{tvla1,tvla2,invisible-invariants,indexed-predicate-abstraction,reps-vmcai}. Some
of these, following Yorsh, \emph{et al.}~\cite{yorsh} use sampling
with a decision procedure to evaluate the abstract
transformer~\cite{thakur-cav12}. Interpolation has also
been used to compile efficient symbolic abstract
transformers~\cite{jhala-mcmillan}. However, these techniques are restricted to
finite domains or domains of finite height  to allow convergence.
Here, we use infinite parameterized domains to obtain better
generalization; hence, the abstract transformer computation is more
challenging. Nonetheless, the approach might also be applicable in
verification.

\vspace{-8pt}
\section{Limitations}

\vspace{-5pt}
While this paper takes a first step towards automatically inferring useful abstractions for synthesis, our proposed method has the following limitations:

\vspace{-5pt}
\paragraph{Shapes of transformers.} Following prior work~\cite{blaze}, our algorithm assumes that abstract transformers have the shape given in \eqnref{trans}. We additionally assume that  constants $\vec{c}$ used in  predicate templates are numeric values and that  functions in \eqnref{trans} are affine. This assumption holds in several domains considered in prior work~\cite{morpheus,blaze} and allows us to develop an efficient learning algorithm that reduces the problem to solving a system of linear equations. %While it is possible to develop more general transformer learning algorithms (e.g., using CEGIS~\cite{cegis1,cegis2}),  

\vspace{-5pt}
\paragraph{DSL semantics.} Our method requires the DSL designer to provide the DSL's logical semantics. We believe that giving logical semantics is much easier than coming up with useful abstractions, as it does not require insights about the internal workings of the synthesizer. Furthermore, our technique could, in principle, also work without logical specifications although the learnt abstract domain may not be as effective (see Footnote 3 in Section~\ref{sec:abstract}) and the synthesized transformers would not be provably sound. 
%Furthermore, we could, in principle, just use the c (rather than axiomatic) semantics 

\vspace{-5pt}
\paragraph{UBT assumption.} Our completeness and termination theorems are predicated on the \emph{unique best transformer (UBT)} assumption. While this assumption holds in our evaluation, it may not hold in general. However, as mentioned in Section~\ref{sec:theorems}, we can always guarantee termination by including the concrete predicates used in the interpolant $\itp$ in addition to the symbolic templates extracted from $\itp$.
\vspace{-8pt}
\section{Conclusion}

\vspace{-5pt}
We  proposed a new technique for automatically instantiating abstraction-guided synthesis frameworks in new domains. Given a DSL and a few training problems, our method automatically discovers a useful abstract domain and the corresponding transformers for each DSL construct.  From a technical perspective, our method uses tree interpolation to extract reusable templates from failed synthesis attempts and automatically synthesizes unique best transformers if they exist. We have incorporated the proposed approach into the \blaze meta-synthesizer and show that the abstractions discovered by \atlas are very useful.

While we have applied the proposed technique to program synthesis, we believe that some of the ideas introduced here are more broadly applicable. For instance, the idea of extracting reusable predicate templates from interpolants and synthesizing transformers in a data-driven way could also be useful in the context of program verification.

%\bibliography{main}

%\appendix

%\input{proofs}

\end{document}